\documentclass[%
 reprint,
superscriptaddress,
 amsmath,amssymb,
 aps,
]{revtex4-2}

\usepackage{graphicx}
\usepackage{dcolumn}
\usepackage{bm}
\usepackage{color}
\usepackage{amsfonts}
\usepackage{dsfont}
\usepackage{amsmath}
\usepackage{amssymb}

\usepackage[
]{changes}

\usepackage{verbatim}
\usepackage{booktabs}
\usepackage{framed}
\usepackage[T1]{fontenc}
\usepackage[utf8]{inputenc}
\usepackage{graphicx}
\usepackage{wrapfig}
\usepackage{appendix}
\usepackage{listings}
\usepackage{xcolor}
    \definecolor{purple}{RGB}{128,0,128}
\usepackage{sidecap}
\usepackage{braket}
\usepackage[hidelinks]{hyperref}
\usepackage{tikz}
\usetikzlibrary{positioning}

\usepackage{pifont}

\newcommand{\D}{\text{d}}
\newcommand{\eye}{\mathds{1}}
\newcommand{\ten}{\otimes}
\newcommand{\Tr}{\text{Tr}}

\begin{document}


\title{Single-photon nonlocality in quantum networks}

\author{Paolo Abiuso} 
\thanks{These authors contributed equally to this work.}
\affiliation{ICFO – Institut de Ciencies Fotoniques, The Barcelona Institute of Science and Technology, 08860 Castelldefels (Barcelona), Spain}

\author{Tamás Kriváchy} 
\thanks{These authors contributed equally to this work.}
\affiliation{ICFO – Institut de Ciencies Fotoniques, The Barcelona Institute of Science and Technology, 08860 Castelldefels (Barcelona), Spain}
\affiliation{Department of Applied Physics, University of Geneva, CH-1211 Geneva, Switzerland}
\affiliation{Institute for Quantum Optics and Quantum Information — IQOQI Vienna,
Austrian Academy of Sciences, Boltzmanngasse 3, 1090 Vienna, Austria}
\affiliation{Atominstitut, Technische Universit\"at Wien, 1020 Vienna, Austria}

\author{Emanuel-Cristian Boghiu}
\affiliation{ICFO – Institut de Ciencies Fotoniques, The Barcelona Institute of Science and Technology, 08860 Castelldefels (Barcelona), Spain}

\author{Marc-Olivier Renou}
\affiliation{ICFO – Institut de Ciencies Fotoniques, The Barcelona Institute of Science and Technology, 08860 Castelldefels (Barcelona), Spain}

\author{Alejandro Pozas-Kerstjens}
\affiliation{Departamento de Análisis Matemático, Universidad Complutense de Madrid, 28040 Madrid, Spain}
\affiliation{Instituto de Ciencias Matem\'aticas (CSIC-UAM-UC3M-UCM), Madrid, Spain}

\author{Antonio Ac\'in}
\affiliation{ICFO – Institut de Ciencies Fotoniques, The Barcelona Institute of Science and Technology, 08860 Castelldefels (Barcelona), Spain}
\affiliation{ICREA - Instituci\'o Catalana de Recerca i Estudis Avan\c{c}ats, Lluis Companys 23, 08010 Barcelona, Spain}

\date{\today}

\begin{abstract}
A single-photon maximally entangled state is obtained when a photon impinges on a balanced beamsplitter. Its nonlocal properties have been intensively debated in the quantum optics and foundations communities.
It is however clear that a standard Bell test made only of passive optical elements cannot reveal the nonlocality of this state. 
We show that the nonlocality of single-photon entangled states can nevertheless be revealed in a quantum network made only of beamsplitters and photodetectors.
In our protocol, three single-photon entangled states are distributed in a triangle network, introducing indeterminacy in the photons' paths and creating nonlocal correlations without the need for measurements choices.
We discuss a concrete experimental realisation and provide numerical evidence of the tolerance of our protocol to standard noise sources. 
Our results show that single-photon entanglement may constitute a promising solution to generate  genuine network-nonlocal correlations useful for Bell-based quantum information protocols.
\end{abstract}

\maketitle

\section{Background}

Local hidden variables models cannot account
for all the predictions of quantum theory. This was 
formalized in 1964 by J.~S. Bell \cite{bell1964einstein}, and is now commonly termed nonlocality \cite{brunner2014bell}. 
Nonlocality is a quantum property with no classical analogue displayed in the so-called Bell tests, defined by the statistics obtained when performing appropriate local measurements on a well-chosen entangled state. Bell tests have been performed in many different systems, from massive particles~\cite{Hanson2015} to photons~\cite{Zeilinger15,NIST15}, and using many different degrees of freedom, such as electronic levels, polarization, orbital angular momentum or time bins. 
In most of these realizations the relevant degrees of freedom used to encode the entanglement are transmitted to each distant observer by a physical carrier, such as, for instance, a photon.

In this work we are interested in the question of whether single-particle quantum states can display nonlocal correlations with no classical analogue. In particular, we consider the question in the context of single-photon entanglement, that is, the state
\begin{equation}\label{eq:SinglePhotonMaxEntState}
\ket{\psi^+}_{AB}=\frac{1}{\sqrt{2}}(\ket{01}_{AB}+\ket{10}_{AB}),
\end{equation}
obtained when sending a single photon into a balanced beamsplitter. Here $\ket{01}_{AB}$ (resp. $\ket{10}_{AB}$) represents the situation in which the photon is sent to the right party $B$ (resp. the left party $A$). The resulting state therefore consists of only one photon and entanglement is encoded in the two optical spatial modes. 

Is the state~\eqref{eq:SinglePhotonMaxEntState} nonlocal? This question has been intensively debated in the quantum foundations and quantum optics community, e.g.~\cite{TWC91,Hardy94,Gerry96,vaidman1995nonlocality,AV2000,Bjork2004,Bellini2006,Brask2013,Sangouard2013,Wamsley2014,zukowski2021, yurke1992bell,yurke1992einstein,gebhart2021genuine}. In principle, a positive answer is provided by the following simple argument~\cite{Gerry96,vaidman1995nonlocality,AV2000}: the two optical modes can be transferred to the population of two energy levels of two distant massive particles. Single-photon entanglement is therefore mapped into two-particle entanglement and a Bell test can now be implemented. The question is much subtler when considering only optical means. To obtain a nonlocal behavior, the two observers need to use local \emph{active measurements} involving local oscillators creating extra local photons~\cite{TWC91,Hardy94,Brask2013,zukowski2021}: without these active measurements, measuring the information content of the state~\eqref{eq:SinglePhotonMaxEntState} would allow the observers to deduce if they received the photon sent by the source, destroying the indeterminacy in the photon path, i.e. the coherences in~\eqref{eq:SinglePhotonMaxEntState}. Then, the statistics become classically simulable. One is therefore tempted to conclude that the observation of nonlocal effects in the single-photon entangled state by passive optical means, that is, phase shifters, beamsplitters and photodetectors, is impossible.

The main result of this work is to show that this is not the case and one can indeed reveal the nonlocality of state~\eqref{eq:SinglePhotonMaxEntState} with only \emph{passive measurements}. To do so, we go beyond standard Bell tests and consider setups defined by causal networks.
These are causal structures involving several independent sources, each being distributed to a subset of the parties involved in the scenario, according to a structure defined by a network~\cite{networkreview}.
It is well understood that these networks offer new possibilities to design quantum experiments with no classical analogue~\cite{Branciard2010,Fritz2012,Fritz2016,Chaves2017,VanHimbeeck2019,renou2019genuine}.
Here, we show that three copies of single-photon entangled states placed in a \emph{triangle causal network} (cf. Fig.~\ref{fig:setup}) can exhibit non-classical correlations. 
Our main idea is to exploit the topology of the network to reintroduce indeterminacy in the photon path, necessary to exploit the coherences of these states.
Remarkably, the obtained setup is not only passive in terms of the implemented measurements, but also because it does not require any active choice of measurements. That is, in our setup, there are no classical inputs and observers perform a single measurement on their received shares. These characteristics make the proposal, arguably, the simplest experimental demonstration of the nonlocality of the single-photon entangled state, as well as the first experimental proposal for genuine network nonlocality~\cite{renou2019genuine}.

Beyond the fundamental motivation, our results are also relevant from an applied point of view. Correlations with no classical analogue are the main resource for device-independent applications. For instance, the security of device-independent protocols for quantum random number generation~\cite{ColbeckPhD,Pironio2010} and quantum key distribution~\cite{Acin2007} is based on the observation of Bell inequality violations.
For that, the simplest way of producing entangled states is through Spontaneous Parametric Down Conversion (SPDC). Entanglement can be encoded on different degrees of freedom of the resulting two photons. However, the state produced by SPDC is a mixture of the desired entangled state and vacuum~\cite{Sangouard2015}. In fact, a heralded preparation of a two-photon maximally entangled state is quite challenging~\cite{Banaszek2003}. In turn, single-photon entanglement can be easily prepared in a heralded way: an arbitrarily good approximation to it can be obtained when detecting photons in one of the two modes resulting from the SPDC process and sending the non-measured mode into a balanced beamsplitter (cf.~\cite{supmat}). Moreover, this form of entanglement does not require the control of any other light degrees of freedom, such as, e.g., polarization or orbital angular momentum. Therefore, the design of simple setups to generate correlations with no classical analogue from this state opens new avenues for the implementation of device-independent protocols.

\section{The triangle network}

\begin{figure}
\centering
\includegraphics[width=0.5\textwidth]{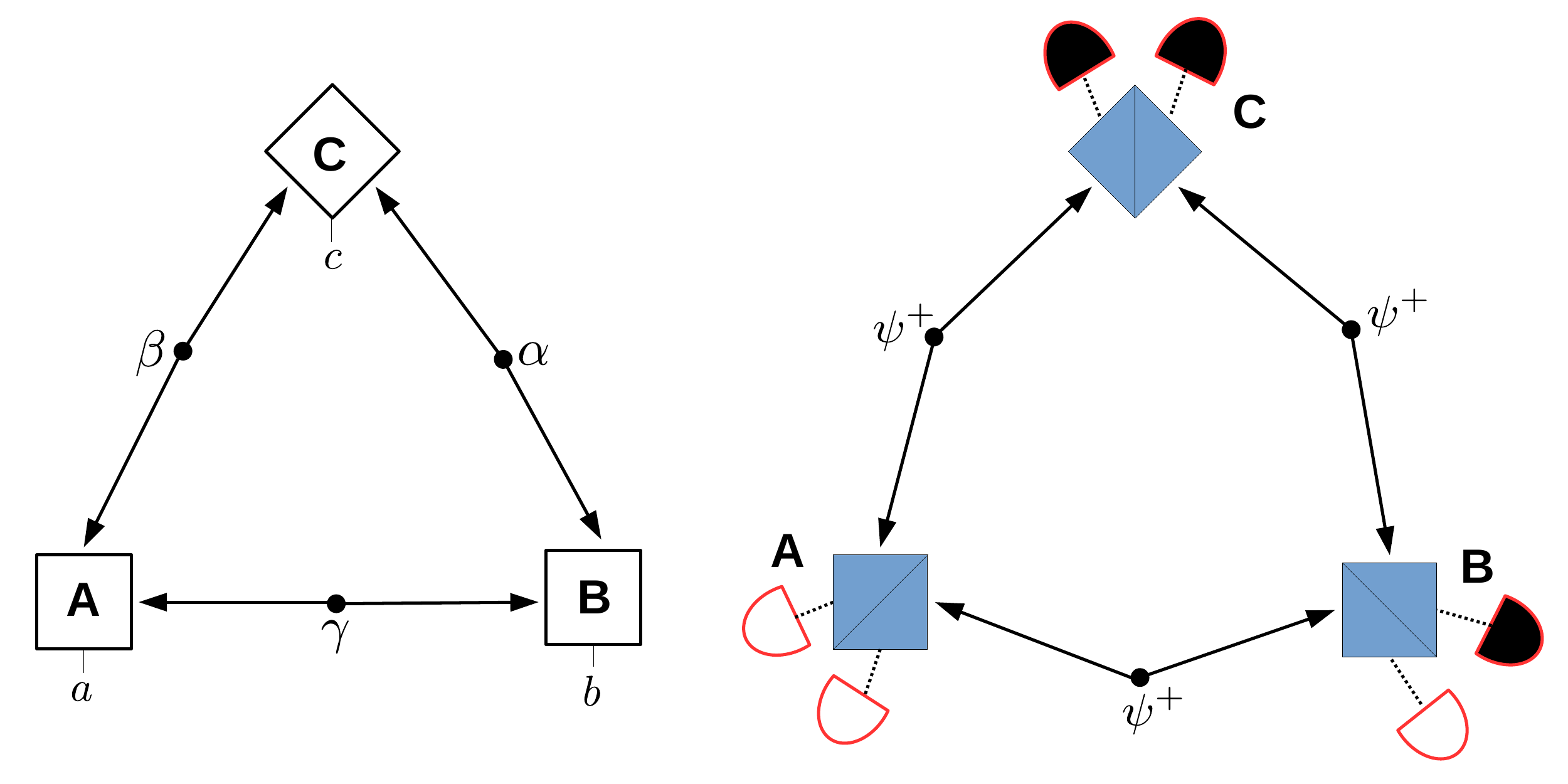}
\caption{(Left) Causal model for the Triangle Network: three independent sources $\{\alpha,\beta,\gamma\}$ prepare correlated states that are distributed among the three parties. Each of them produces an output through a local process acting on the received parts of the states. The form of the states and local processes depend on the theory, say classical or quantum, used to reproduce the correlations in the network. (Right)~Schematics of the proposed quantum optical experiment. A, B and C share single-photon entangled states $\ket{\psi^+}=(\ket{01}+\ket{10})/\sqrt{2}$ prepared by the sources. Each party receives two optical modes that are mixed on a beamsplitter, the resulting output modes being measured by photodetectors. In the specific experimental instance depicted here, A does not detect any photon, B has one detector firing, and C has both detectors firing.}
\label{fig:setup}
\end{figure}

The considered Bell-type experiment consists of a triangle causal network where three observers, $A$, $B$ and $C$, receive states prepared by three sources, see Fig.~\ref{fig:setup}. These states are measured producing outcomes $a$, $b$ and $c$ with probability $p(abc)$.

\label{sec:ideal_exp}

A classical description of the experiment compatible with the causal constraints defined by the network has the form (here $\D\alpha$, $\D\beta$ and $\D\gamma$ are normalized measures)
\begin{align}
p(abc)=\int \D\alpha \D\beta \D\gamma\;  p_A(a|\beta\gamma) p_B(b|\gamma\alpha) p_C(c|\alpha\beta)\;.
\label{eq:TriangleDef}
\end{align}
The causal model therefore consists of classical variables $\alpha$, $\beta$ and $\gamma$ distributed by the sources and local response functions $p_X$, with $X=A,B,C$, producing the measurement outcomes. In analogy with standard Bell tests, we define probability distribution $p(abc)$ that can be written as Eq.~\eqref{eq:TriangleDef} as causally classical or, simpler, \emph{local}.

A quantum description of the experiment compatible with the causal network replaces the random variables by quantum states $\rho_\alpha$, $\rho_\beta$ and $\rho_\gamma$ and the local response functions by quantum measurements. Therefore, quantum probabilities compatible with the triangle network have the form
\begin{align}
p(abc)=\Tr\left[(\rho_\alpha \otimes \rho_\beta \otimes \rho_\gamma) (M^{(a)}_A \otimes M^{(b)}_B \otimes M^{(c)}_C)\right]\;,
\label{eq:QTriangleDef}
\end{align}
where $M^{(a)}_A$ denote the positive measurement operators defining the Positive-Operator Valued Measure (POVM) for $A$, $\sum_a M^{(a)}_A=\eye_A$, and similarly for B and C. We slightly abuse the notation in Eq.~\eqref{eq:QTriangleDef} by not specifying the tensor products and different Hilbert spaces in which the different operators act, but this is clear from Fig.~\ref{fig:setup}. We say that a quantum experiment, defined by states and measurements producing the outcome distribution $p(abc)$ according to Eq.~\eqref{eq:QTriangleDef}, is \emph{nonlocal} whenever this distribution cannot be described by a classical model~\eqref{eq:TriangleDef}. Our goal in what follows is to provide a nonlocal quantum experiment in the triangle network using only single-photon entangled states, beamsplitters and photodetectors.

The basic idea of the experimental proposal is depicted in Fig.~\ref{fig:setup}: three parties $A$, $B$, $C$ share, for each pair $AB$, $BC$, $CA$, the single photon entangled state $|\psi^+\rangle$, see Eq.~\eqref{eq:SinglePhotonMaxEntState}. The initial state is thus
\begin{equation}
|\psi^+\rangle_{A_2B_1}\ten|\psi^+\rangle_{B_2C_1}\ten|\psi^+\rangle_{C_2A_1}\equiv \ket{\Psi^+}_{A_1A_2B_1B_2C_1C_2}\ .
\end{equation}
Each party then receives its two optical inputs on modes $X_1 X_2$ ($X=A,B,C$) and mixes them with a beamsplitter, which induces a unitary transformation $\mathcal{B}_{X_1X_2}(t,\phi)$ parametrized by its transmissivity $t$ and phase $\phi$.
All parties use the same value for $t$, and the phases are all null for simplicity in the following (cf.~\cite{supmat}).

After passing through the beamsplitters, the photons end up in photodetectors. For each mode $X_i$, the operators describing a perfectly efficient photodetection correspond to the projectors onto the vacuum state $D^{\square}_{X_i}=|0\rangle\langle 0|_{X_i}$ (detector off) and the projector on its orthogonal complement $D^{\blacksquare}_{X_i}=\eye_{X_i}-|0\rangle\langle 0|_{X_i}$ (detector firing). Indeed, we assume that the detectors do not resolve the number of photons but only their presence.
The measurement obtained by mixing two modes with the beamsplitter and the ideal photodetectors can be accordingly expressed as a POVM for each party (here $\mathcal{B}_{X_1X_2}=\mathcal{B}_{X_1X_2}(t,0)$)
\begin{align}
\nonumber
    {\Pi_t^{(0)}}_{X_1X_2}&=\mathcal{B}^\dagger_{X_1X_2}(D^{\square}_{X_1}\ten D^{\square}_{X_2})\mathcal{B}_{X_1X_2},\\
\nonumber
    {\Pi_t^{(L)}}_{X_1X_2}&=\mathcal{B}^\dagger_{X_1X_2}(D^{\blacksquare}_{X_1}\ten D^{\square}_{X_2})\mathcal{B}_{X_1X_2}, \\
    {\Pi_t^{(R)}}_{X_1X_2}&=\mathcal{B}^\dagger_{X_1X_2}(D^{\square}_{X_1}\ten D^{\blacksquare}_{X_2})\mathcal{B}_{X_1X_2},\\
    \nonumber
    {\Pi_t^{(2)}}_{X_1X_2}&=\mathcal{B}^\dagger_{X_1X_2}(D^{\blacksquare}_{X_1}\ten D^{\blacksquare}_{X_2})\mathcal{B}_{X_1X_2},
    \label{eq:POVM_main}
\end{align}
where the measurement labels stand respectively for no photon counts ($0$), a count in the left detector ($L$), a count in the right detector ($R$), or counts in both detectors ($2$). The crucial point is that when $t\neq 0$, the $L$ and $R$ measurements actually detect superpositions of photons in the incoming modes (see details in~\cite{supmat}).

The quantum experiment described here results in the output distribution
\begin{align}
\nonumber
 p_t(abc) &=\Tr[\ket{\Psi^+}\!\bra{\Psi^+}
 ({\Pi_t^{(a)}}
 \otimes{\Pi_t^{(b)}}
 \otimes{\Pi_t^{(c)}}
 )]
  \\ 
  a,b,c &\in \{0,L,R,2\}
\end{align} 
which depends on the transmissivity $t$ of the beamsplitters used by the parties and whose exact expression can be found in the Supplementary Material~\cite{supmat}.

\section{Witnessing single-photon nonlocality}
\label{sec:witnessing}
The first main result of this work is that
\begin{flushleft}
\emph{The distribution $p_t$ obtained from the experiment described in Fig.~\ref{fig:setup} (cf. previous section), is nonlocal (at least) for values of the beamsplitter transmissivity in the intervals $t \in (0,0.215)$ and $t \in (0.785,1)$.}
\end{flushleft}

We give in the following a sketch of the proof, which is analytical and detailed in~\cite{supmat}. 

First, we simplified the structure that classical strategies must follow in the triangle network~\eqref{eq:TriangleDef}.
Specifically, all the local response functions $p_A$, $p_B$, $p_C$ in~\eqref{eq:TriangleDef} can be assumed to be deterministic, and all the indeterminacy is therefore delegated to the classical sources $\{\alpha,\beta,\gamma\}$, which can all be assumed to be, w.l.o.g, real numbers uniformly distributed in the interval $[0,1]$. Therefore, any local model is specified by deterministic triangle-local response functions $p_A p_B p_C$ that map all the points of the cube $[0,1]^3$ to the observed outputs 
\begin{align}
\label{eq:response_map}
    \{\alpha,\beta,\gamma\}\rightarrow\{a(\beta,\gamma),b(\gamma,\alpha),c(\alpha,\beta)\}\;.
\end{align}

Secondly, we were able to identify strict constraints that need to be satisfied by all possible classical causal models simulating the considered experimental output $p_t(abc)$ in the triangle network. In particular, we exploited the cyclic symmetry and null components 
of the distribution.
For example, all outputs of the form (here $\chi$ represents any of $L$ or $R$) 
$\{(000), (00\chi), (2\chi\chi), (22\chi)\}$, or any of their permutations,
have zero probability, due to the fact that there are initially 3 photons in the network, of which at most 2 can end up in the same photodetector. That is, in each run of the experiment the total number of clicks in the detectors must be 2 or 3. 
By taking all the relevant properties of $p_t$ into account, one can identify constraints that need to be satisfied by any classical strategy, specified by the response functions~\eqref{eq:response_map}, aiming at reproducing $p_t$. 
In fact, while the exact form of the response functions remains in general unknown, some of its marginals can be expressed in terms of the output $p_t$. These relevant marginals are nothing other than linear constraints on the response functions, parametrized by $t$. Together with standard normalization and positivity constraints, these define a Linear Program.
The feasibility of such Linear Program is, by definition, necessary for the existence of such local response functions. Therefore, when infeasible, no local model exists to simulate our experiment proposal. Results show that the Linear Program is infeasible for $t \in (0.785,1)$ and $t \in (0,0.215)$, proving the claims of this section. We refer to the Supplementary Material for the technical details and the complete proof.

The techniques we used are similar to those  introduced in~\cite{renou2019genuine} and generalized in~\cite{renou2020network}. However, their findings cannot be applied directly to our scenario. 
The reason behind this is that the works~\cite{renou2019genuine,renou2020network} are based on a token-counting approach to some physical "tokens" that are: $i)$~generated from the sources, $ii)$~distributed to the parties in a coherent superposition of different ways, and $iii)$~counted at the output. 
In our experiment the physical tokens are the photons, which however can be miscounted at the output, as more than one could enter in the same photodetector.
For these reasons, in the proof~\cite{supmat} we had to extend these techniques so that they could be applied to our setup.
As part of the proof, we showed that our distribution is nonlocal if and only if the distribution proposed in~\cite{renou2019genuine}, which we dub $p'_t$, is nonlocal as well. 
While finishing this manuscript, we became aware of preliminary unpublished results~\cite{InPrep}, which prove nonlocality of $p'_t$ for discrete points in the range $t \in (0.5,0.785)$ as well. 
Nonlocality of $p'_t$ in such interval has been conjectured already~\cite{krivachy2020neural}. Given the above mentioned equivalence between the nonlocality of $p_t$ and $p'_t$ proven in this work, this would imply that the proposed ideal experiment is nonlocal for all transmissivities except  $t\in\{0.0,0.215,0.5,0.785,1.0\}$, which are known to have local models (cf.~\cite{renou2019genuine,supmat}).

\section{Noise tolerance and machine learning analysis}

After proving the nonlocality of the outputs of the ideal noiseless experiment, we analyzed the robustness of our results against typical noise errors, by modelling imperfections which occur in experimental realizations of the optical network presented in Fig.~\ref{fig:setup}. Therefore, the resulting output distribution, $p^{Q,T,\nu}_t(abc)$ depends on additional noise parameters quantifying: the impurity of the generated single-photon entangled state ($Q$), the transmissivity of the optical channels ($T$) of the network, and the efficiency of the final photodetectors ($\nu$). It follows that
\begin{align}
  p^{Q=0,T=1,\nu=1}_t(abc)  \equiv p_t(abc)\;,
\end{align}
that is, with no impurity, and perfect transmission and detection, we recover the idealized experiment. The details of the modelling employed are deferred to the Sup.Mat.~\cite{supmat}.

Inevitably, part of the key properties and symmetries of $p_t(abc)$ disappear as soon as noise is introduced in the network. This makes the analytic approach unworkable in this case.
Consequently, in order to estimate the tolerance to the noises introduced above, we resorted to a  technique recently introduced in \cite{krivachy2020neural}: there, a feed-forward neural network is shaped with the same topology of the causal network under study, and it is then asked to reproduce the target distribution $p_t^{(Q,T,\nu)}$. Each output of the neural network is thus \emph{literally} an instance of a classical model (which can be therefore described by Eq.~\eqref{eq:TriangleDef} in our case) trying to reproduce $p_t^{(Q,T,\nu)}$. For a fixed target distribution, the neural network is trained by minimizing the Euclidean distance from the neural network's local model to the target. When the target distribution is inside the local set, a sufficiently large neural network should be capable of learning it. Instead, a large distance between the machine's best guess and the target is taken as an indication of nonlocality. What it means to be ``large'' enough can be somewhat arbitrary, since some nonlocal behaviors are extremely close to the local set (as is the case here), and additionally the neural network's model is not guaranteed to converge to the optimal solution as it can get stuck in local minima during training. In order to gain deeper insight into the boundary between locality and nonlocality we examine transitions of the learning algorithm's behavior when adding noise to the target distribution, and retraining the machine independently for each target distribution. The very noisy case is guaranteed to be local and the machine learning results on those give a reference to which we can compare the nonlocal regime. By definition, this technique does not certify nonlocality in an absolute way, but has been shown to be reliable and efficient from the point of view of computational resources \cite{krivachy2020neural}.

\begin{figure}
\centering
\includegraphics[width=0.49\textwidth]{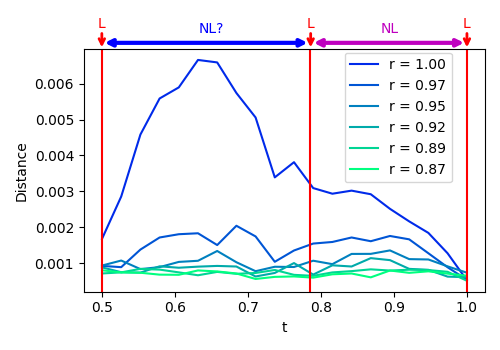}
\caption{Euclidean distance of machine learned local models to the target distributions $p_t(abc)$, for various levels of artificial noise on the singlets~\eqref{eq:SinglePhotonMaxEntState} (visibilities $r$ of Werner states $r\ket{\psi^+}\bra{\psi^+} + (1-r)\eye/4$). With red vertical lines we depict the transmissivities $t$ at which analytic local models exist  ($t\in\{0.5,0.785,1\}$). At the top of the figure a purple line shows the regime where we have proven nonlocality, while the blue line shows the regime where we conjecture nonlocality, based on these numerics and the relation to the distribution in Ref.~\cite{renou2019genuine}, which was studied numerically in Ref.~\cite{krivachy2020neural}.}
\label{fig:ml_results_1}
\end{figure}

\begin{figure}[t!]
\centering
\includegraphics[width=0.49\textwidth]{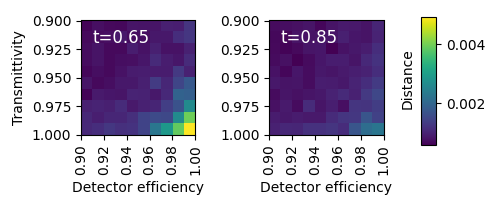}
\caption{Euclidean distance of machine learned local models from the noisy distribution $p_t^{(Q,T,\nu)}$ under an experimentally realistic noise model for $t=0.65$ (left) and $t=0.85$ (right), with $Q=0.006875$ for both.
}
\label{fig:ml_results_2}
\end{figure}

The results of the analysis are summarized in Figs.~\ref{fig:ml_results_1} and~\ref{fig:ml_results_2}, where we consider only $t\geq 0.5$ because of the symmetry of the experiment when mirroring the beamsplitters $t'=1-t$. 
For the noiseless distribution (perfect visibility $r=1$ in Fig.~\ref{fig:ml_results_1}), the neural network's best guess is distant from the experimental output, corroborating the analytical proof of nonlocality for $t \in (0.785,1)$. At the same time the neural network hints at the locality of the output distribution for $t=0.5$ and $t=1$, which clearly have local strategies. A local model exists as well for $t\sim 0.785$ (cf.~\cite{renou2019genuine,supmat}) where the neural network struggles to get closer; however, note that the distance of ~0.003 achieved there is already \emph{very} close to the local set.
Moreover, the same machine indicates (seemingly even stronger) nonlocality in the range
$t \in (0.5,0.785)$, in line with the conjecture of~\cite{krivachy2020neural} and the results of~\cite{InPrep}.

The noise robustness is, however, small. In Fig.~\ref{fig:ml_results_1} an artificial noise is considered by adding a Werner state visibility to the source~\eqref{eq:SinglePhotonMaxEntState} of ideal experiment $(Q=0,T=1,\nu=1)$.
The neural network seems to indicate that the points that are ``most nonlocal'' are $t\sim 0.85$ in the proven region (purple interval in Fig.~\ref{fig:ml_results_1}), and $t\sim 0.65$ in the conjectured region (blue interval). For these two points we tested the tolerance to the physical noises introduced above, see Fig.~\ref{fig:ml_results_2}: choosing $Q\simeq 0,7 \%$ (cf.~\cite{supmat}), the neural network tries to learn $p_t^{(Q,T,\nu)}$ for different values of the transmissivity $T$ and detector efficiency $\nu$. Results show that nonlocality is more robust for $t=0.65$, where it is lost when $T \lesssim 95\%$ or $\nu\lesssim 95\%$.

All data was obtained by representing each of the three response function ($p_A(a|\gamma\beta), p_B(b|\gamma\alpha), p_C(c|\alpha\beta)$) by a multilayer perceptron of depth 4 and width 20 with rectified linear activation functions. For each target distribution we retrained the neural network independently 30 times and kept the smallest distance among those.

\section{Discussion}
We have proven how single-photon entangled states can be used to generate an outcome distribution with no classical analogue in a triangle network. The considered setup only requires passive optical elements, namely beamsplitters, phase shifters and photodetectors, and involves a single measurement per observer. Our results not only challenge the current understanding of the nonlocal properties of single-photon entanglement, but also open new perspective for the use of this form of entanglement for quantum information applications, as they provide the first proposal of an experimental demonstration of genuine network nonlocality.

We have shown that the nonlocality of such proposal has (small) noise-tolerance to natural noises that can arise in its implementation, through a machine learning analysis. Such approach is however not exact, and it remains an open question to prove nonlocality in the noisy regime by other means, e.g. certifying it by inflation techniques \cite{wolfe2019inflation}, which would be crucial for an experimental implementation.

Finally, in the Supplementary Material~\cite{supmat} we show that our main result on the nonlocality of the ideal experimental proposal in the triangle network can be extended to any ring network with $N\geq 3$ parties, although increasing the number of parties does not improve the detectability of nonlocality in the proposed experiment with our current techniques.

\begin{acknowledgements}
The authors thank Mattieu Perrenoud, Nicolas Maring, Nicolas Brunner, Nicolas Gisin for insightful discussions.
This work is supported by the Government of Spain (FIS2020-TRANQI and Severo Ochoa CEX2019-000910-S), Fundacio Cellex, Fundacio Mir-Puig, Generalitat de Catalunya (CERCA, AGAUR SGR 1381 and QuantumCAT), the ERC AdG CERQUTE, the AXA Chair in Quantum Information Science.
P. A. is supported by ``la Caixa" Foundation (ID 100010434, Grant  No.  LCF/BQ/DI19/11730023).
T. K. is supported by the Swiss National Science Foundation (Starting grant DIAQ, QSIT, Doc.Mobility), and the European Research Council (ERC MEC).
E.-C. B. received funding from the “Presidencia de la Agencia Estatal de Investigación” (Ref. PRE2019-088482). 
M.-O. R. is supported by the Swiss National Fund Early Mobility Grant P2GEP2\_191444.
A. P.-K. is supported by the European Union's Horizon 2020 research and innovation programme-grant agreement No. 648913 and by the Spanish Ministry of Science and Innovation through the ``Severo Ochoa Programme for Centers of Excellence in R\&D'' (CEX2019-000904-S).
The machine learning computations were performed at University of Geneva on the ``Baobab'' HPC cluster.
\end{acknowledgements}

\medskip
\bibliography{BIB.bib}


\pagebreak
\ 
\newpage
\widetext
\begin{center}
\textbf{\large Supplemental Material for ``Quantum networks reveal single-photon nonlocality''}
\end{center}
\setcounter{equation}{0}
\setcounter{figure}{0}
\setcounter{table}{0}
\setcounter{page}{1}
\setcounter{section}{0}
\makeatletter
\renewcommand{\theequation}{S\arabic{equation}}
\renewcommand{\thefigure}{S\arabic{figure}}

\section{Noiseless output distribution}
\label{app:noiseless_dist}

Here we derive the form of the noiseless output distribution $p^{Q=0,T=1,\nu=1}_t(abc)  \equiv p_t(abc)$ produced when all the elements of the optical scheme described in Sec. \ref{sec:ideal_exp} are perfect.

The initial state shared among the parties is 
\begin{align}
    \ket{\psi^+}\!\bra{\psi^+}_{A_2B_1}\ten \ket{\psi^+}\!\bra{\psi^+}_{B_2C_1}\ten\ket{\psi^+}\!\bra{\psi^+}_{C_2A_1}\;, \quad \text{with}\quad \ket{\psi^+}=\frac{\ket{01}+\ket{10}}{\sqrt{2}}\;.
\end{align}

The action of a beamsplitter with transmissivity $t$ and phase $\phi$ is described in terms of the input and output optical modes with creation operators $a^\dagger_i$ as
\begin{equation}
\label{eq:beam_splitter}
\begin{pmatrix}
a^{\dag}_2 \\
a^{\dag}_1
\end{pmatrix}_{in}
=
\begin{pmatrix}
\sqrt{t} & -e^{-i\phi}\sqrt{1-t} \\
e^{i\phi}\sqrt{1-t} & \sqrt{t}
\end{pmatrix}
\begin{pmatrix}
a^{\dag}_2 \\
a^{\dag}_1
\end{pmatrix}_{out}\;.
\end{equation}
Consequently, the corresponding unitary induced by the transformation can be derived in the Fock basis by expressing $|mn\rangle_{X_2X_1}\equiv\frac{a^{\dag m}_2}{\sqrt{m!}}\frac{a^{\dag n}_1}{\sqrt{n!}}|00\rangle_{X_2X_1}$,
 to obtain
\begin{align}
|10\rangle_{in} =& \sqrt{t}|10\rangle_{out}-e^{-i\phi}\sqrt{1-t}|01\rangle_{out}\;, \\
|01\rangle_{in} =& \sqrt{t}|01\rangle_{out}+e^{i\phi}\sqrt{1-t}|10\rangle_{out}\;, \\
|00\rangle_{in} =& |00\rangle_{out}\;, \\
|11\rangle_{in} =& (2t-1)|11\rangle_{out}-e^{-i\phi}\sqrt{2t(1-t)}|02\rangle_{out} +e^{i\phi}\sqrt{2t(1-t)}|20\rangle_{out}  \;.
\end{align}
Accordingly, the POVM~\eqref{eq:POVM_main} can be written as
\begin{align}
\label{eq:povm00}
 \Pi^{(0)}_t=&  |00\rangle\langle 00|\;,\\
 \label{eq:povm01}
 \Pi^{(R)}_t=&  |\chi_r\rangle\langle \chi_r|+2t(1-t)|11\rangle\langle 11|\;,
\\
\label{eq:povm10}
\Pi^{(L)}_t=&   |\chi_l\rangle\langle \chi_l|+2t(1-t)|11\rangle\langle 11|\;,
\\
\label{eq:povm11}
 \Pi^{(2)}_t=&   (2t-1)^2|11\rangle\langle 11|\;.
\end{align}
where $|\chi_r\rangle=\sqrt{t}|01\rangle-e^{i\phi}\sqrt{1-t}|10\rangle$, $|\chi_l\rangle=\sqrt{t}|10\rangle+e^{-i\phi}\sqrt{1-t}|01\rangle$, and where we truncated the Hilbert space considering that the input state consists only of combinations of vacuum and a single-photon excitation. 
Therefore each party has four possible outputs $a,b,c\in\{0,L,R,2\}$, standing for no detector counts $\square\square$, a count in the left detector ${\blacksquare}\square$, a count in the right detector $\square {\blacksquare}$, or counts in both detectors $\blacksquare\blacksquare$, respectively, described by the POVM above.

The resulting network output 
\begin{align}
 p_t(abc) &=\Tr[\big(\ket{\psi^+}\!\bra{\psi^+}_{A_2B_1}\ten \ket{\psi^+}\!\bra{\psi^+}_{B_2C_1}\ten\ket{\psi^+}\!\bra{\psi^+}_{C_2A_1}\big)\big({\Pi_t^{(a)}}_{A_1A_2}\otimes{\Pi_t^{(b)}}_{B_1B_2}\otimes{\Pi_t^{(c)}}_{C_1C_2}\big)]
\end{align} 
has multiple constraints due to the cyclic symmetry of the experiment, due to all the parties using the same value for the beamsplitter transmissivity $t$ \eqref{eq:beam_splitter}, as well as photon number conservation. For example, all outputs of the form (here $\chi$ represents any of $L$ or $R$)
\begin{align}
\label{eq:photon_cons}
p_t(000)=0\;, \quad & p_t(00\chi)=0\;, \quad & \text{(too few photons would be detected)} \\
p_t(2\chi\chi)=0\;, \quad & p_t(22\chi)=0\;, & \quad \text{(too many photons would be detected)}
\end{align}
are null, due to the fact that there are initially 3 photons in the network, of which at most 2 can end up in the same photodetector.

The non-zero probabilities are, modulo the cyclic symmetry, in the form $p(0\chi\chi)$, $p(02\chi)$, $p(0\chi 2)$, $p(\chi\chi\chi)$, and are summarised, in order, in the following.
\begin{align}
 p_t(0LL)=p_t(0RR)=\frac{1}{4}t(1-t), \quad p_t(0RL)=\frac{1}{2}t(1-t)^2, \quad p_t(0LR)=\frac{1}{2}t^2(1-t),
 \end{align}
 \begin{align}
 p_t(02R)=& \frac{1}{8}(2t-1)^2 t, \quad & p_t(02L)=& \frac{1}{8}(2t-1)^2(1-t),\\
 p_t(0R2)=& \frac{1}{8}(2t-1)^2(1-t), \quad & p_t(0L2)=& \frac{1}{8}(2t-1)^2 t,
 \end{align}
 \begin{align}
\nonumber
 p_t(RRL)=& \frac{1}{8}t(1-t)(1+2\cos(\Phi)\sqrt{t(1-t)}),\quad & p_t(LLR)=& \frac{1}{8}t(1-t)(1-2\cos(\Phi)\sqrt{t(1-t)}),\\ 
 p_t(LLL)=& \frac{1}{8}(1-3t(1-t)+2t^{\frac{3}{2}}(1-t)^{\frac{3}{2}}\cos(\Phi)), \quad & p_t(RRR)=& \frac{1}{8}(1-3t(1-t)-2t^{\frac{3}{2}}(1-t)^{\frac{3}{2}}\cos(\Phi)).
\end{align}
where $\Phi\equiv\phi_A+\phi_B+\phi_C$.
 
In what follows, we take $\Phi=0$, as the range of values of $t$ for which the distribution is proven to be nonlocal decreases when $\Phi\neq 0$ (that is, the following analysis can be performed for an arbitrary value of $\Phi$, and the interval of values of $t$ for which $p_t$ is nonlocal is maximised when $\Phi=0$).
Also, note that $\Phi=\phi_A+\phi_B+\phi_C$ can be tuned locally by any of the parties.

\section{Nonlocality of the noiseless distribution}
\label{app:analytics_details}

To prove the nonlocality of the ideal noiseless distribution $p_t$ presented above, we take an approach inspired by the one presented in \cite{renou2019genuine}.
There, a quantum distribution is proposed, which is based on the
same input state in the triangle network (we report it in our notation)
\begin{equation}
\label{eq:sourceapp}
|\psi^+\rangle_{A_2B_1}\ten|\psi^+\rangle_{B_2C_1}\ten|\psi^+\rangle_{C_2A_1}\equiv \ket{\Psi^+}_{A_1A_2B_1B_2C_1C_2}
\end{equation}
with $\ket{\psi^+}=\frac{\ket{01}+\ket{10}}{\sqrt{2}}$,
and the following POVM on the two modes $X_2X_1$ of each party $X=A,B,C$ (again, we use a notation that makes the comparison easier with the experiment proposed in the present manuscript)
\begin{align}
\Pi'^{(0)}_t=|00\rangle\langle00|, \quad \Pi'^{(R)}_t=|\chi_r \rangle\langle \chi_r|, \quad \Pi'^{(L)}_t=|\chi_l \rangle\langle \chi_l|, \quad \Pi'^{(2)}_t=|11\rangle\langle 11|,
\end{align}
where $|\chi_r\rangle=\sqrt{t}|01\rangle-\sqrt{1-t}|10\rangle$ and $|\chi_l\rangle=\sqrt{t}|10\rangle+\sqrt{1-t}|01\rangle$ (here we put all the phases $\phi_x$ to zero, as mentioned above).
The output distribution  of our experiment is not equivalent to that of \cite{renou2019genuine}, as our POVM consists, as described in Sec.~\ref{app:noiseless_dist}, of
\begin{align}
\nonumber
  \Pi^{(0)}_t &=|00\rangle\langle 00|, \quad 
 \Pi^{(R)}_t =|\chi_r\rangle\langle \chi_r|+2t(1-t)|11\rangle\langle 11|, \\
\Pi^{(L)}_t &=|\chi_l\rangle\langle \chi_l|+2t(1-t)|11\rangle\langle 11|, \quad
 \Pi^{(2)}_t =(2t-1)^2|11\rangle\langle 11|.
\end{align}
Notice that both POVMs $\Pi$ and $\Pi'$ are a coarse graining of the measurement
\begin{align}
\nonumber
  \Pi''^{(0)}_t=|00\rangle\langle 00|, \quad 
 \Pi''^{(R1)}_t=|\chi_r\rangle\langle \chi_r|,\quad \Pi''^{(R2)}_t=2t(1-t)|11\rangle\langle 11|, \\
\Pi''^{(L)}_t=|\chi_l\rangle\langle \chi_l|,\quad \Pi''^{(L2)}_t=2t(1-t)|11\rangle\langle 11|, \quad
 \Pi''^{(2)}_t=(2t-1)^2|11\rangle\langle 11|.
\end{align}
The POVM $\Pi''$ is the one that would be obtained from the scheme described in the main text if the photodetectors were able to resolve photon numbers, and has thus six possible outputs (cf.\eqref{eq:povm00}-\eqref{eq:povm11}).
Accordingly, it is possible to define distributions $p_t$, $p'_t$, $p''_t$,  obtained from the state~\eqref{eq:sourceapp} and applying (respectively) $\Pi_t$, $\Pi'_t$, $\Pi''_t$ at each party modes $X_2X_1$, i.e.
\begin{align}
    p_t(abc) &=\Tr[\ket{\Psi^+}\!\bra{\Psi^+}_{A_1A_2B_1B_2C_1C_2}({\Pi_t^{(a)}}_{A_1A_2}\otimes{\Pi_t^{(b)}}_{B_1B_2}\otimes{\Pi_t^{(c)}}_{C_1C_2})] \\
    \label{eq:p'_def}
    p'_t(abc) &=\Tr[\ket{\Psi^+}\!\bra{\Psi^+}_{A_1A_2B_1B_2C_1C_2}({\Pi_t'^{(a)}}_{A_1A_2}\otimes{\Pi_t'^{(b)}}_{B_1B_2}\otimes{\Pi_t'^{(c)}}_{C_1C_2})] \\
    p''_t(abc) &=\Tr[\ket{\Psi^+}\!\bra{\Psi^+}_{A_1A_2B_1B_2C_1C_2}({\Pi_t''^{(a)}}_{A_1A_2}\otimes{\Pi_t''^{(b)}}_{B_1B_2}\otimes{\Pi_t''^{(c)}}_{C_1C_2})]
\end{align}

Surprisingly, we prove that $p_t$, $p'_t$, and $p''_t$,
have the same range of nonlocality for the parameter $t$. That is, for a fixed $t$, if one among $p_t$, $p'_t$, $p''_t$, is classically reproducible in the triangle network, then all of them are. At the same time, the infeasibility of one among $p_t$, $p'_t$, $p''_t$, implies the infeasibility of all of them. From the physical point of view, this means that the possibility of performing perfect number-resolving photodetection does not enhance the ``nonlocality'' of the output distribution of our ideal experiment, although it may improve its resistance to noise.

To prove the nonlocal equivalence (in the triangle network) of the three distributions $p_t$, $p'_t$, $p''_t$ we proceed as follows:
\begin{align}
\label{eq:feasibility_implications}
    \text{feasibility $p'_t$} \Rightarrow
    \text{feasibility $p''_t$} \Rightarrow
    \text{feasibility $p_t$} \Rightarrow
    \text{feasibility $p'_t$} \;,
\end{align}
where by ``feasibility'' we mean the feasibility of classically simulating the distribution with a local model, as from Eq.~\eqref{eq:TriangleDef}. The first two implications follow immediately, without assumptions on the input state $\ket{\Psi}$, from simple properties of the POVMs involved. Indeed:
\begin{itemize}
    \item The POVM $\Pi''$ can be obtained as a fine-graining of $\Pi'$ via a probabilistic splitting of $\Pi'^{(2)}$ in three outcomes $\Pi''^{(L2)}$, $\Pi''^{(R2)}$, $\Pi''^{(2)}$, which is just a classical local post-processing of the original projector $\ket{11}\!\bra{11}$.
    \item The POVM $\Pi$ is a local coarse-graining of $\Pi''$ and thus $p_t$ is classically simulatable whenever $p''_t$ is. 
\end{itemize}
The last implication requires more effort and we prove it in the following subsections. To do so, we identify constraints on local strategies simulating $p_t$ and show that these are the same as those needed to simulate $p'_t$, as from~\cite{renou2019genuine} (cf. following derivations and Paragraph~\ref{subsec:equivalence_pp'}).

\subsection{Constraints on local models simulating $p_t$}

We start by assuming that there exists a classical model that simulates the output distribution $p_t$ of the experiment proposed in the main text, and we find the constraints that it has to respect. That is, we assume that indeed $p_t$ (which is summarised in Sec.~\ref{app:noiseless_dist}), can be written as
\begin{align}
p_t(abc)=\int_0^1 \int_0^1 \int_0^1 \D\alpha \D\beta \D\gamma\; p_A(a|\beta\gamma) p_B(b|\gamma\alpha) p_C(c|\alpha\beta)\ .
\end{align}
Notice that the classical shared variables $\{\alpha,\beta,\gamma\}$ can be assumed to be real numbers in the $[0,1]$ interval, and all the randomness of the local statistical responses $p_X$ can be absorbed in the distribution of $\{\alpha,\beta,\gamma\}$, meaning that without loss of generality the local response functions can be taken deterministic, i.e.,
\begin{align}
p_A(a|\beta,\gamma)=\delta(a-A(\beta,\gamma)).
\end{align}
where $A(\beta,\gamma)$ is some deterministic response function.
Let $X$, $Y$ and $Z$ denote the set of possible $\alpha$, $\beta$ and $\gamma$ respectively. Let us define
\begin{align}
\nonumber
 X_{0}^B = \{\alpha\, |\, \exists\, \gamma: B(\gamma,\alpha) = 0 \}
\qquad&
 X_{2}^B = \{\alpha\, |\, \exists\, \gamma: B(\gamma,\alpha) = 2 \}
\\
\nonumber
 X_{2}^C = \{\alpha\, |\, \exists\, \beta: C(\alpha, \beta) = 2 \}
\qquad&
 X_{0}^C = \{\alpha\, |\, \exists\, \beta: C(\alpha, \beta) = 0 \}
\\
\nonumber
 Y_{0}^C = \{\beta\, |\, \exists\, \alpha: C(\alpha,\beta) = 0 \}
\qquad&
 Y_{2}^C = \{\beta\, |\, \exists\, \alpha: C(\alpha,\beta) = 2 \}
\\
\nonumber
 Y_{2}^A = \{\beta\, |\, \exists\, \gamma: A(\beta, \gamma) = 2 \}
\qquad&
 Y_{0}^A = \{\beta\, |\, \exists\, \gamma: A(\beta, \gamma) = 0 \}
\\
\nonumber
 Z_{0}^A = \{\gamma\, |\, \exists\, \beta: A(\beta, \gamma) = 0 \}
\qquad&
 Z_{2}^A = \{\gamma\, |\, \exists\, \beta: A(\beta, \gamma) = 2 \}
\\
 Z_{2}^B = \{\gamma\, |\, \exists\, \alpha: B(\gamma, \alpha) = 2 \}
\qquad&
 Z_{0}^B = \{\gamma\, |\, \exists\, \alpha: B(\gamma, \alpha) = 0 \}
 \label{eq:sets_def}
\end{align}
In short, set $X_{i}^P$ is the set of $\alpha$'s for which party $P$ can potentially obtain output $i$, and similarly for the $Y_i^P$ and $Z_i^P$ sets for $\beta$'s and $\gamma$'s, respectively.

We coarse-grain the possible outcomes by grouping outcomes $L$ and $R$ as $\chi$, which means that the possible outcomes are now $a,b,c \in \{0,\chi, 2\}$. Then, according to Sec.~\ref{app:noiseless_dist}, the set of outcomes with nonzero probability in our setup are (up to permutations)
\begin{align}
\label{eq:outcomes}
abc \in \{\chi \chi \chi, 0\chi \chi, 0\chi2\}.
\end{align}
Observe that
\begin{itemize}
\item two 0's never appear at the same time, nor two 2's,
\item 2 only appears together with exactly one $\chi$ and one 0.
\end{itemize}
These properties are simply due to the fact that the number of photons is conserved, and that at most two photons can end up in the same photodetector. Already from these observations we obtain some structure on the previously defined sets in three steps. We demonstrate the steps for the $\{X_i^P\}_{i,P}$ sets, but they can be done with the $\{Y_i^P\}_{i,P}$ and $\{Z_i^P\}_{i,P}$ similarly.
\subsubsection{$X_2^B \cap X_2^C = \emptyset$, $X_0^B \cap X_0^C = \emptyset$}
\label{subsec:const1}
This is a direct consequence of the previous observation (Eq.~\ref{eq:outcomes}). There cannot be 4 photons among 4 parties or 0 photons in total for two parties.
\subsubsection{$X_0^B \cup X_0^C = X$}
\label{subsec:const2}
Assume by contradiction that $\exists\ \alpha^* \in X\setminus \left( X_0^B \cup X_0^C \right)$. Then by definition $\forall \beta,\gamma:$
\begin{align*}
 B(\gamma, \alpha^*) \in \{\chi,2\},\\
 C(\alpha^*,\beta) \in \{\chi,2\}.
\end{align*}
Observe that when $\alpha=\alpha^*$, Alice must not answer $a=2$, due to (\ref{eq:outcomes}). However, to do this, since she does not know the value of $\alpha$, Alice must \emph{always} not answer $a=2$. A similar conclusion can be drawn for the other parties, due to the cyclic symmetry. This, however, leads to a contradiction since parties can in general output $2$, e.g. $p_t(a=2)\neq 0$. 
\subsubsection{$X_0^B \cap X_2^B = \emptyset$, $X_0^C \cap X_2^C = \emptyset$}
\label{subsec:const3}
Assume by contradiction that $\exists \alpha^* \in X_0^B \cap X_2^B$. Then $\exists \gamma_1, \gamma_2$ s.t.
\begin{align*}
B(\gamma_1, \alpha^*) = 0,\\
B(\gamma_2, \alpha^*) = 2.
\end{align*}
Charlie does not know $\gamma$, so if $\alpha=\alpha^*$, he knows he must answer $\chi$ for \emph{any} $\beta$, since that is the only symbol consistent with both $0$ and $2$. Thus we have that
\begin{align*}
\forall \beta: \;\; C(\alpha^*, \beta) = \chi.
\end{align*}
Say Alice receives $\gamma = \gamma_2$. Alice does not know whether $\alpha = \alpha^*$ or not. Thus, her response must be one that is consistent with the scenario that $\alpha = \alpha^*$. Because of Charlie's response being $\chi$, this implies that for any $\beta$ she must answer $a=0$, i.e.
\begin{align*}
\forall \beta: \;\; A(\beta, \gamma_2) = 0.
\end{align*}
This means, by definition, that $Y_0^A = Y$. This implies, after doing steps 1 and 2 for the sets $\{Y_i^P\}_{i,P}$, that $Y_0^C = \emptyset$. However, since $p_t(c=0)\neq 0$, we arrive at a contradiction.
\begin{figure}
    \centering
    \includegraphics[width = 0.4 \textwidth]{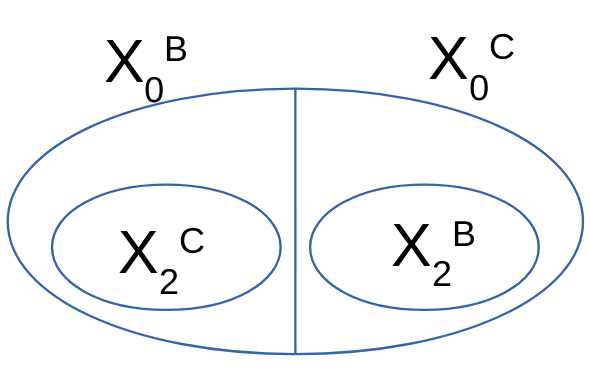}
    \caption{The relation of the sets $\{X_i^P\}_{i,P}$ to each other.}
    \label{fig:sets}
\end{figure}

\subsubsection{All sets $X_0^P,\;Y_0^P,\;Z_0^P$ have probability $1/2$}
\label{sec:all_sets_0.5}
The previous constraints \ref{subsec:const1}-\ref{subsec:const3} on the sets $X_i^P$ are can be summarized as in Fig.~\ref{fig:sets}. We now give a partial quantitative assessment on the size of these sets.
Note that by the definition of the sets $Y_0^A$ and $Z_0^A$ we have
\begin{align}
\label{eq:one-fourth}
\frac{1}{4} = p_t(a=0)  \leq p(\beta \in Y_0^A, \gamma \in Z^A_0) =  p(Y_0^A) p(Z_0^A),
\end{align}
 where in the last step we used the statistical independence of the hidden variables. At the same time, by using the inequality $ab\leq ((a+b)/2)^2$ we have
\begin{align*}
p(Y_0^A) p(Y_0^C) \leq \left(\frac{p(Y_0^A) + p(Y_0^C)}{2}\right)^2 = \frac{1}{4}.
\end{align*}
Combining the two we see that $p(Y_0^C) \leq p(Z_0^A).$ Repeating the same argument (cyclically) for the other parties we get
\begin{align*}
p(Y_0^C) \leq p(Z_0^A) \leq p(X_0^B) \leq p(Y_0^C),
\end{align*}
which implies that they are all equal. Using also (\ref{eq:one-fourth}) it is clear that all sets with $i=0$ are equally probable with probability $\frac{1}{2}$, i.e.,
\begin{align}
\label{eq:all_1/2}
p(Y_0^C) = p(Z_0^A) = p(X_0^B) = p(Y_0^C) 
= p(Y_0^A) = p(Z_0^B) = p(X_0^C) = p(Y_0^A) = \frac{1}{2}.
\end{align}

\begin{figure}
    \centering
	\begin{tikzpicture}[every node/.style={minimum size=1cm},on grid]
	\node at (-1.1,-0.5) {\textcolor{blue}{Face B}};
	\node at (2,4.0) {\textcolor{blue}{Face A}};
	\node at (5.0,-0.5) {\textcolor{blue}{Face C}};
	\begin{scope}[every node/.append style={yslant=-0.5},yslant=-0.5]
	  \shade[right color=gray!10, left color=black!50] (0,0) rectangle +(2,2);
	  \node at (-0.5,1.5) {$X_0^C$};
	  \node at (-0.5,0.5) {$X_0^B$};
	  \node at (0.5,1.5) {$2,\chi$};
	  \node at (1.5,1.5) {$\chi$};
	  \node at (0.5,0.5) {$\chi$};
	  \node at (1.5,0.5) {0};
	  \draw (0,0) grid (2,2);
	\end{scope}
	\begin{scope}[every node/.append style={yslant=0.5},yslant=0.5]
	  \shade[right color=gray!70,left color=gray!10] (2,-2) rectangle +(2,2);
	  \node at (2.5,-0.5) {0};
	  \node at (3.5,-0.5) {$\chi$};
	  \node at (2.5,-1.5) {$\chi$};
	  \node at (3.5,-1.5) {$2,\chi$};
	  \draw (2,-2) grid (4,0);
	\end{scope}
	\begin{scope}[every node/.append style={
	    yslant=0.5,xslant=-1},yslant=0.5,xslant=-1
	  ]
	  \shade[bottom color=gray!10, top color=black!80] (4,2) rectangle +(-2,-2);
	  \node at (4.5,1.5) {$Z_0^A$};
	  \node at (4.5,0.5) {$Z_0^B$};
	  \node at (2.5,2.5) {$Y_0^C$};
	  \node at (3.5,2.5) {$Y_0^A$};
	  \node at (2.5,1.5) {$\chi$};
	  \node at (2.5,0.5) {$2,\chi$};
	  \node at (3.5,1.5) {0};
	  \node at (3.5,0.5) {$\chi$};
	  \draw (2,0) grid (4,2);
	\end{scope}
	\end{tikzpicture}
	\quad
	\begin{tikzpicture}[every node/.style={minimum size=1cm},on grid]
	\begin{scope}
	  \node at (0,-0.5) {}; 
	  \shade[right color=gray!10, left color=black!50] (0,0) rectangle +(2,2);
	  \node at (1,3) {$Z_0^A$};
	  \node at (1.5,2.3) {$Z_2^B$};
	  \node at (-1,1) {$X_0^C$};
	  \node at (-0.3,0.5) {$X_2^B$};
	  \node at (0.5,1.5) {$\chi$};
	  \node at (1.5,1.5) {$\chi$};
	  \node at (0.5,0.5) {$\chi$};
	  \node at (1.5,0.5) {$2,\chi$};
	  \draw (0,0) grid (2,2);
	\end{scope}
	\end{tikzpicture}
	\caption{(Left) Classical strategies visualized on a cube. The edges of the cube represent the interval $[0,1]$, on which the hidden variables $\{\alpha,\beta,\gamma\}$ distributed, and the sets \eqref{eq:sets_def} are represented, all having probability $1/2$ due to \eqref{eq:all_1/2}. The labels on faces are the possible responses of a given party.  (Right) The $2,\chi$ part on face B, for example, can be further decomposed using $X_2^B$ and $Z_2^B$.}
	\label{fig:cube}
\end{figure}
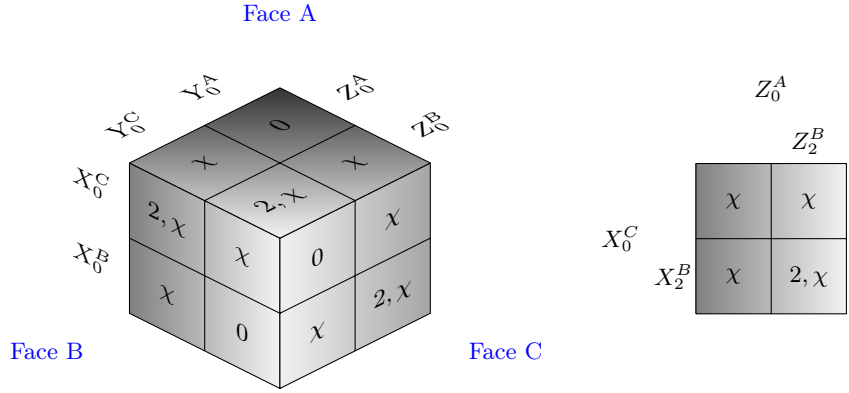

Equation~\eqref{eq:all_1/2} combined with \eqref{eq:one-fourth} tells us that Alice, when receiving from $Y^A_0$ on one side and from $Z^A_0$ on the other, will \emph{deterministically} output $0$. The same holds for the other parties (Bob when receiving from $X^B_0$ and $Z^B_0$, and Charlie when receiving from $X^B_0$ and $Y^B_0$). This consideration combined with the previous ones and the definition of the sets~\eqref{eq:sets_def}, yields a constrained picture of all possible classical models that simulate the coarse graining of $p_t$ in the triangle network. This is illustrated in Fig.~\ref{fig:cube}:
\begin{itemize}
    \item Alice outputs $0$ when receiving from $Y^A_0$ and $Z^A_0$. 
    \item Alice outputs $\chi$ when receiving from $Y^A_0$ and $Z^B_0$, or when receiving from $Y^C_0$ and $Z^A_0$ (in both cases Alice cannot output $a=2$ because of property \ref{subsec:const3}). 
    \item Alice outputs either $\chi$ or $2$ when receiving from $Y^C_0$ and $Z^B_0$,  (further structure can be given using the sets $Z^B_2$ and $X_2^B$).
\end{itemize}
Bob and Charlie follow similar strategies when cycling the indices. 

\subsection{Breaking up the coarse-graining}
\label{sec:breaking_up}

\begin{framed}
\textbf{Now, the main idea is the following:} If there exists a local model for $p_t(a,b,c)$ as from Fig.~\ref{fig:cube}, then there should exist a distribution $q_t(i,j,k,s)$ representing the parties collective response function ($i,j,k=L,R$) when the hidden variables $\alpha,\beta,\gamma$ come from $X_0^C \times Y_0^A \times Z_0^B$ (s=0) or $S_1 = X_0^B \times Y_0^C \times Z_0^A$ (s=1). We cannot directly derive $q_t(i,j,k,s)$ from $p_t(a,b,c)$, however, we can derive its marginals (see below). These marginals will be incompatible for some values of transmissivity $t$. For these situations, thus, we can deduce that there does not exist a local model for $p_t(a,b,c)$. Additionally, the marginals constraints on $q_t(i,j,k,s)$, are the same as in \cite{renou2019genuine} for the distribution $p'_t$, meaning that the classical feasibility of $p_t$ implies the classical feasibility of $p'_t$, as stated in~\eqref{eq:feasibility_implications}.
\end{framed}

To start, consider the two sets $S_0 = X_0^C \times Y_0^A \times Z_0^B$ and $S_1 = X_0^B \times Y_0^C \times Z_0^A$. Note that $S_0 \cap S_1 = \emptyset$ and the events $\chi \chi \chi$ can happen if and only if $(\alpha,\beta,\gamma)\in S_0 \cup S_1$.  We define
\begin{align}
q_t(i,j,k,s) = p(a=i,b=j,c=k, (\alpha,\beta,\gamma) \in S_s\, |\, (\alpha,\beta,\gamma) \in S_0 \cup S_1),
\label{eq:q_DEF}
\end{align}
where the indices $i,j,k$ are each either $L$ or $R$, and the index $s$ is either $0$ or $1$. This is a probability distribution, since if $(\alpha,\beta,\gamma) \in S_0 \cup S_1$, then it must be either in $S_0$ or $S_1$, and all parties must output either $L$ or $R$ (hence normalization and positivity are satisfied). Using the definition of conditional probability and the fact that the sets $S_0$ and $S_1$ have probability $1/8$ (cf. \ref{sec:all_sets_0.5} and Fig.~\ref{fig:cube}),  we see that 
\begin{align}
q_t(i,j,k,s) = 4 p(a=i,b=j,c=k, (\alpha,\beta,\gamma) \in S_s ).
\end{align}
Marginalizing over $s$ gives us
\begin{align}\label{eq:marginal_t}
\boxed{
q_t(i,j,k) = 4 p_t(a=i,b=j,c=k),
}
\end{align}
the value of which is given by the parameters of the model, e.g. the transmissivity.

Next we would like to express other marginals, e.g. $q_t(i,s)\equiv \sum_{jk}q_t(i,j,k,s)$, as a function of the target probability distribution. To do this, first note that if $b=0$ then $\alpha \in Z_0^B, \gamma \in X_0^B$ and \emph{either} $\beta \in Y_0^A$ \emph{or} $\beta \in Y_0^C$. Note that next to a $0$ output we can only have the other two parties answering $\{\chi, 2\}$ or $\{\chi, \chi\}$. For $q_t(i,s)$ we are, however, interested in the probabilities of $a= i$, therefore we break up the $\chi$ in Alice's response. In terms of probabilities this means
\begin{align}
\nonumber
p_t(a=i,\textcolor{blue}{b=0},c=\chi) +& p_t(a=i,\textcolor{blue}{b=0},c=2)
=\\
= p_t(a=i,(\alpha,\beta,\gamma) \in \textcolor{blue}{X_0^B} \times \textcolor{red}{Y_0^A} \times \textcolor{blue}{Z_0^B}) +& p_t(a=i,(\alpha,\beta,\gamma) \in \textcolor{blue}{X_0^B} \times \textcolor{red}{Y_0^C} \times \textcolor{blue}{Z_0^B})
\label{eq:sums}
\end{align}
where we used colors to simplify the reading, separating the sets in a local strategy on which Bob bases his choice (in blue), from those to which he has no access (in red).
From now on we use a shorthand for expressions like this, indicating e.g.  $(\alpha,\beta,\gamma) \in {X_0^B} \times {Y_0^C} \times {Z_0^B}$ simply as ${X_0^B} {Y_0^C} {Z_0^B}$. 

Next, consider the sum where we force Alice to output $i$, but Bob and Charlie can either output $0$ or $\chi$. In other words we are focusing on the $\chi 0 \chi, \chi 0 2, \chi \chi 0, \chi 2 0$ outputs, breaking coarse-graining $\chi\rightarrow L,R$ only in Alice's case. Define the quantity $D_A^i$ as
\begin{align}
D_A^i:=  p_t(a=i,\textcolor{blue}{b=0},c=\chi)
 +p_t(a=i,\textcolor{blue}{b=0},c=2)
&-\left[ p_t(a=i,b=\chi,\textcolor{blue}{c=0})
 +p_t(a=i,b=2,\textcolor{blue}{c=0}\right]. \end{align}
 A few manipulations show that 
 \begin{align}
 \nonumber
D_A^i=
  p_t(a=i,\textcolor{blue}{X_0^B} \textcolor{red}{Y_0^A} \textcolor{blue}{Z_0^B}) 
 +p_t(a=i,\textcolor{blue}{X_0^B} \textcolor{red}{Y_0^C} \textcolor{blue}{Z_0^B})
&-\left[ p_t(a=i,\textcolor{blue}{X_0^C Y_0^C} \textcolor{red}{Z_0^A}) 
 +p_t(a=i,\textcolor{blue}{X_0^C Y_0^C} \textcolor{red}{Z_0^B})\right] =  \\
=
  p_t(a=i,\underline{\mathbf{X_0^C} Y_0^A Z_0^B})
 +p_t(a=i,\mathbf{X_0^C} Y_0^C Z_0^B)
&-p_t(a=i,\underline{\mathbf{X_0^B} Y_0^C Z_0^A}) 
 -p_t(a=i,X_0^C Y_0^C Z_0^B)=\\
=
p_t(a=i,\underline{S_0})
&- p_t(a=i,\underline{S_1}),
\end{align}
where we first used (\ref{eq:sums}) (and a similar expression for $c=0$), and then that Alice does not have access to $\alpha$, so the probabilities stay the same under the swap of $X_0^B$ for $\mathbf{X_0^C}$ and $X_0^C$ for $\mathbf{X_0^B}$. Finally we identified ${S_0}$ and ${S_1}$ in the relevant expressions.
Hence, we could express the differences of $q(i,s=0)$ and $q(i,s=1)$ as an expression of known terms. We also know that the sum is
\begin{align}
p_t(a=i,S_0) + p_t(a=i,S_1) = p(a=i, b=\chi,c=\chi) = \sum_{j,k=L,R}p(a=i, b=j,c=k).
\end{align}
Combining the two we get that
\begin{align}
\boxed{q(i,s=0) =  2p_t(a=i, b=\chi,c=\chi) + 2D_A^i}\\
\boxed{q(i,s=1) =  2p_t(a=i, b=\chi,c=\chi) - 2D_A^i}
\end{align}

\subsection{Testing $q_t(i,j,k,s)$ using linear programming}
We sum up here the marginal properties (boxed equations in the previous section \ref{sec:breaking_up}) of the distribution $q_t(i,j,k,s)$ found above. These properties are linear constraints on the vector $q_t(i,j,k,s)$ which are parametrized by the transmissivity $t$. A  linear program can be implemented to verify if a distribution a $q_t(i,j,k,s)$  compatible with these marginals exists. 

\paragraph{Constraint 0 (normalization).} First of all,
\begin{equation}
q_t(i,j,k,s)\geq 0 \ \forall i,j,k,s  \qquad \text{and} \qquad \sum_{i,j,k,s} q_t(i,j,k,s)=1 
\end{equation}
that is, it truly represents a probability vector.

\paragraph{Constraint 1.} 
Then
\begin{equation}
\sum_s q_t(i,j,k,s)=4p_t(a=i,b=j,c=k)
\end{equation}
with (cf. Sec.~\ref{app:noiseless_dist})
\begin{align}
\nonumber
 4p_t(RRL)=\frac{1}{2}t(1-t)(1+2\sqrt{t(1-t)}),\quad 4p_t(RLL)=\frac{1}{2}t(1-t)(1-2\sqrt{t(1-t)}), \\
 4p_t(LLL)=\frac{1}{2}((1-t)^{\frac{3}{2}}+t^{\frac{3}{2}})^2, \quad 4p_t(RRR)=\frac{1}{2}((1-t)^{\frac{3}{2}}-t^{\frac{3}{2}})^2.
\end{align}
and cyclic combinations (meaning only the number of $L$s and $R$s matters).

\paragraph{Constraint 2.}
This constraint is actually a consequence of Constraint 1, but we write it for completeness.
\begin{equation}
\sum_{s,j,k}q_t(i,j,k,s)=4p_t(a=i, \chi,\chi) \quad \text{and cyclic cases } i\rightarrow j\rightarrow k\rightarrow i \;.
\end{equation}

\paragraph{Constraint 3.}
\begin{equation}
\sum_{j,k}q_t(i,j,k,s=0)-q_t(i,j,k,s=1)=4 \big[ p_t(a=i,0,\chi)+p(a=i,0,2)-p(a=i,\chi,0)-p(a=i,2,0)\big] 
\end{equation}
and cyclic combinations.
This last constraint can be made explicit (cf. Sec.~\ref{app:noiseless_dist})
\begin{align}
& i=L \rightarrow 4 \big[ p_t(a=i,0,\chi)+p_t(a=i,0,2)-p_t(a=i,\chi,0)-p_t(a=i,2,0)\big]=\frac{1}{2}-t \;,\\
& i=R \rightarrow 4 \big[ p_t(a=i,0,\chi)+p_t(a=i,0,2)-p_t(a=i,\chi,0)-p_t(a=i,2,0)\big]=t-\frac{1}{2} \;.
\end{align}
\paragraph{Relation to Ref.\cite{renou2019genuine} and equivalence between $p_t$ and $p'_t$}
\label{subsec:equivalence_pp'}
The constraints defining the linear program above, can be translated to be the same constraints of a linear program found in Ref.~\cite{renou2019genuine}, where the distribution $p'_t$~\eqref{eq:p'_def} is considered (in~\cite{renou2019genuine} $t$ is identified as $u^2$). Specifically, both the distributions $p_t$ and $p'_t$ are local if a solution $\bar{q}_t(i,j,k,t)$ to the same linear program exists and can be generated via a local model (cf.~\cite{renou2019genuine}). This proves that the local feasibility of $p_t$ is equivalent to that of $p'_t$. At the same time, the existence of $\bar{q}_t(i,j,k,t)$ is a \emph{necessary condition} for the local feasibility of $p_t$. This means that when the linear program fails to find a solution, the nonlocality of $p_t$ is certified, while if a solution is found, this does not directly imply the locality of $p_t$.

The Linear Program resulting from the constraint above is infeasible for $t \in (0,0.215)$  and $t \in (0.785,1)$.

\section{Noisy optical realisation}
\label{sec:imperfections}
As introduced in the main material, after proving the nonlocality of the idealized experiment, in this section we give the modelling details of the imperfections that can arise in the different elements of the optical network presented in Fig.~\ref{fig:setup}, when realized experimentally. We focused on: 

\emph{a.} the impurity of the generated single-photon entangled state ($Q$), 

\emph{b.} the transmissivity of the optical channels ($T$) of the network, and 

\emph{c.} the efficiency of the final photodetectors ($\nu$). 

Our results (see Main Material) indicate that the noise tolerance w.r.t. these parameters is of the order of few percentage points, which makes the proposal very stringent from the experimental point of view, but possible on a table-optical experiment with high-efficient detectors.

\paragraph{Source imperfections}
\label{app:state_prep}
Firstly, we considered a realistic process of creation for the single photon entangled state $\ket{\psi^+}=(\ket{01}+\ket{10})/\sqrt{2}$. This is generated by a single photon sent onto a 50:50 beamsplitter. Typical  sources achieve the heralding of single photons from two-photon states created in a SPDC process, followed by the detection of one of the two photons~\cite{boyd2020nonlinear,couteau2018spontaneous}.

An externally controlled laser pulses at high frequency on a $\chi^{(2)}$ non-linear crystal. For each pulse, the crystal consequently outputs a two-mode squeezed vacuum state $|\Psi\rangle\propto\sum_n q^n|nn\rangle$. Then, photodetection is performed on one of the two modes. Conditioning on a detection allows to isolate a very good approximation of the one-photon Fock state on the unmeasured mode \cite{christ2012limits}. The trade-off between probability of heralding and quality (fidelity to target) of the heralded state is strongly conditioned by the photodetector efficiency and ability to resolve photon number, as well as the characteristics of the crystal and the laser power, which tune the value of $q$ \cite{christ2012limits}. Here we chose typical currently achievable values for the SPDC, which we assume to have $q=0.01$ and $10$MHz frequency of the pulses \cite{caspar2020heralded}. The heralding is simulated by currently available number-resolving photodetectors which we assume to have 8-photon resolution achieved with an array of $M=8$ single photon detectors pixels, having each a $\eta=70\%$ efficiency, well in the range of present technologies \cite{moshkova2019high,zhu2020resolving}.
Conditioning on the firing of a single pixel in the detector, the resulting state in the unmeasured mode can be approximated by
\begin{align}
\label{eq:single_photon_source}
    \varrho \sim (1-Q)\ket{1}\bra{1}+Q\ket{2}\bra{2}+\mathcal{O}(Q^2)\;,
\end{align}
where $Q\propto q$ is the ratio between the chance of obtaining a single pixel firing due to a double-photon hitting the detector, and the chance of obtaining a single pixel fire due to a single photon, i.e.
\begin{align}
    Q=\dfrac{q^2\left(\frac{1}{M}(1-(1-\eta)^2)+ 2\frac{M-1}{M}\eta(1-\eta)\right)}{q\eta}\ .
\end{align}
Note that the probability of heralding is $~q\eta$ and thus for the three sources (of the experiment proposed in the main text) to be heralded at the same time, the corresponding total experimental repetition rate is of approximately $q^3\eta^3 10\text{MHz}\sim1\text{Hz}$~\footnote{Notice that lasers pulsed at GHz rates have been used recently~\cite{ngah2015ultra}, which would result in an experimental repetition rate of $\sim\text{KHz}$.}.
Considering the imperfect state $\varrho$~\eqref{eq:single_photon_source}, propagated through a 50:50 beamsplitter, the resulting true source shared by each couple in the triangle network is
\begin{align}
    \rho=(1-Q)\ket{\psi^+}\bra{\psi^+}+Q\ket{\varphi}\bra{\varphi}+\mathcal{O}(Q^2)\;.
    \label{eq:source_state}
\end{align}
With the above-mentioned values of $q$, $\eta$, and $M$, it results $Q= 0.006875$. 

Notice that the same single-photon preparation could be done with simple, non-number-resolving (NNR) photodetection. In such a case the value of $Q$ (which we remind, is the ratio between the chance of the detector clicking due to a double-photon, and the chance of a click due to a single-photon), would be
\begin{align}
    Q^{\rm (NNR)}=\frac{q^2(1-(1-\eta)^2)}{q\eta}=q(2-\eta)\;,
\end{align}
where $\eta$ is the efficiency of the detectors. We see that in such a case $Q$ is bounded to be larger than $q$, for example with the same values above ($q=0.01$, $\eta=70\%$), one obtains $Q^{\rm (NNR)}=0.013$, essentially double what can be obtained with number-resolving detectors. This is not a huge limitation per se, as we can rescale $q$ to make $Q^{\rm (NNR)}$ smaller. At the same time, halving $q$ makes the total repetition rate of the experiment ($\propto q^3\eta^3$) decrease by one order of magnitude.

Finally, let us notice how basing our proposal on the single-photon state $\ket{\psi^+}\propto \ket{01}+\ket{10}$ is crucial in our scenario. A unitarily equivalent state is the two-photon state $\propto \ket{HV}+\ket{VH}$, which encodes the information in the polarization degree of freedom. However the creation of such state from an SPDC source typically needs the heralding of the 6-photons term $\ket{33}$ from $\sum_n q^n \ket{nn}$ (and 4 photodetectors per source)~\cite{Banaszek2003}. This means that even in an ideal scenario in which all detectors have unit efficiency, the probability of heralding the correct state would be $\sim q^3$, and for the whole experiment with $3$ sources, $q^9$, compared to $q^3$ for our single-photon proposal. For a $1\%$ error in the source, we chose $q= 0.01$, which is translated into $12$ orders of magnitude of difference in the heralding rate.

\paragraph{Losses in the channels}
Secondly, loss might happen during the transmission along the channels that form the sides of the triangle network of Fig.~\ref{fig:setup}, before the local POVM performed by the parties. We denote by $T$ the transmissivity of these optical channels. 
The resulting correction due to photon loss can be computed as
\begin{align}
\label{eq:st_ph_loss}
    \delta\rho=\sum_{n,X_i}K^{(n)}_{X_i}
    \rho K^{(n)\dagger}_{X_i}
\end{align}
where Kraus operators of the form 
\begin{align}
    K^{(n)}_{X_i}=\sqrt{(1-T)}\sqrt{n}\ket{n-1}_{X_i} \!\bra{n} ,
\end{align}
act on each of the six modes $X_i$, and the sum is truncated to $n=1,2$ (given the support of input state~\eqref{eq:source_state}). In fact, as we work in the regime of low losses, we only keep the first-order terms in $1-T$ in Eq.~\eqref{eq:st_ph_loss}.

\paragraph{Detectors}
Finally, the photodetectors used at the vertices of the triangle (Fig.~\ref{fig:setup}) do not resolve photon number, and are assumed to have a finite, high efficieny $\nu$, thus modelled, at first order in $1-\nu$ as
\begin{align}
\nonumber
   D^{\square}(\nu)=& \ket{0}\bra{0}+(1-\nu)\ket{1}\bra{1}\;, \\ D^{\blacksquare}(\nu)=& \eye -(1-\nu)\ket{1}\bra{1} -\ket{0}\bra{0}\;.
\end{align}
Notice that high efficiencies close to 100\% have been reached by modern photodetection systems \cite{natarajan2012superconducting,lita2008counting,miller2011compact,fukuda2011titanium,reddy2019exceeding}.

\section{Generalization to chains of $N$ parties}
\label{app:ring_net}

\begin{framed}
\textbf{In this section, we sketch a generalization of the experiment presented in the main text (which is proposed in the triangle scenario), to a chain of $N$ parties in a circular network.} For such case, we generalise the procedure carried out through Sec.~\ref{app:analytics_details} which proves the existence of a range of transmissivities for which the network output is nonlocal.
\end{framed}

The generalized experiment is described as follows: $N$ parties $A_i$ share a copy of the single photon state $\ket{\psi^+}=\frac{\ket{01}+\ket{10}}{\sqrt{2}}$ for each couple of neighbouring parties $A_iA_{i+1}$ with $i=1,\dots,N$ (the total network is circular and thus we identify $N+1\equiv 1$). Each party consequently receives two input modes containing at most $1$ photon, and performs the same measurement described in the main text~\eqref{eq:POVM_main}, and detailed in Sec.~\ref{app:noiseless_dist}, consisting in a local mixing of the modes with a beamsplitter of transmissivity $t$, followed by photodetection on both modes. All the parties choose the same value for $t$ and the photodetectors do not resolve the number of photons, thus being described by projective measurements on vacuum and its orthogonal complement $M^{\square}=|0\rangle\langle 0|,\; M^{\blacksquare}=\eye-|0\rangle\langle 0|$.
Consequently, the resulting output distribution is given by
\begin{align}
\label{eq:quantum_circ_p}
    p_t(a_1,\dots,a_n)=\Tr\left[\left(\bigotimes_{i=1}^N \psi^+_{A_i^{(R) }A_{i+1}^{(L)}}\right)\left(\bigotimes_{j=1}^N \Pi^{(a_j)}_{t_{A_j^{(L)}A_j^{(R)}}}\right)\right]\;, \quad a_j=0,L,R,2\;,
\end{align}
where the state $\psi^+\equiv\ket{\psi^+}\bra{\psi^+}$ is shared between each ``right mode'' of the $i$th party ($A_i^{(R)}$) and the ``left mode'' ($A_{i+1}^{(L)}$) of the following, and each party performs the POVM operationally described above, corresponding to $\Pi_t$~\eqref{eq:POVM_main} (detailed in Eq.s~\eqref{eq:povm00}-\eqref{eq:povm11}) on its two modes.

\begin{figure}
\centering
\includegraphics[width=0.7\textwidth]{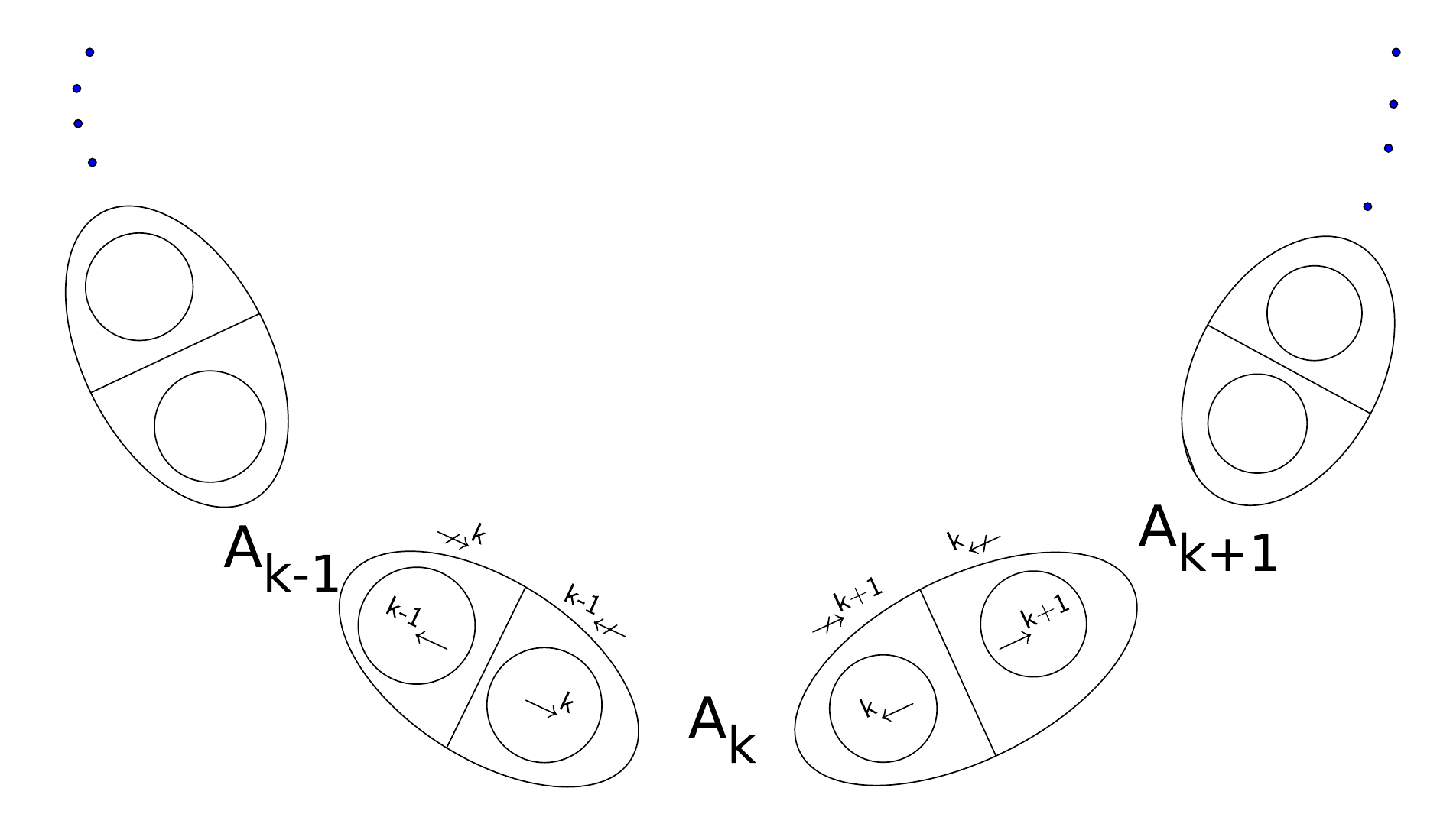}
\caption{Generalized setting with $N$ parties $A_k$, and representation of the sets \eqref{eq:setapp1}-\eqref{eq:setapp4} describing a local strategy that simulates $p_t$ (cf. Eq.s~\eqref{eq:quantum_circ_p} and \eqref{eq:local_circ_model}).}
\label{fig:n-parties}
\end{figure}

We now put constraints on any possible local strategy aiming at reproducing the same statistical output of $p_t$ in the circular network. That is we assume $p_t$ can be written as
\begin{align}
\label{eq:local_circ_model}
p_t(a_1\dots a_N)=\int \D\alpha_{12} \D\alpha_{23}\dots\D\alpha_{N1}\; p_{A_1}(a_1|\alpha_{N1}\alpha_{12}) p_{A_2}(a_2|\alpha_{12}\alpha_{23}) \dots p_{A_N}(a_N|\alpha_{(N-1) N}\alpha_{N1})\ 
\end{align}
where $a_i$ is the output of party $A_i$, which is based on a local response on the hidden variables $\{\alpha_{i(i+1)},\alpha_{(i-1)i}\}$ shared with his left and right neighbours.
In the coarse grained scenario, parties can output $0,\chi,2$ as before ($\chi$ is the coarse graining of $\{L,R\}$, cf. Sec.~\ref{app:analytics_details}), representing the outcomes with $0$, $1$, or $2$ photodetectors firing respectively at each party station.
Following Sec.~\ref{app:analytics_details} we define the equivalent of the sets~\eqref{eq:sets_def}, accompanying the formal definitions with an intuitive notation and explanation of the underlying local model;
the sets are represented by arrows that intuitively suggest the direction of "classical photons" in a corresponding local hidden variable model. The following definitions are pictured in Figure~\ref{fig:n-parties}. We have formally, for the set of sources $\alpha_{(k-1)k}$ between $A_{k-1}$ and $A_k$ ,
\begin{align}
\label{eq:setapp1}
(\nrightarrow^k):= & \{\alpha_{(k-1)k}\, |\, \exists\, \alpha_{k(k+1)}: A_k(\alpha_{(k-1)k},\alpha_{k(k+1)}) = 0 \}\\
\nonumber
\ & \text{This is the set allowing $A_k$ to output $0$ for some of the hidden variables that come from the other side.} \\
\nonumber
& \text{That is, \emph{classical photons are not sent to $A_k$ from the left.}}\\
\label{eq:setapp2}
(^{k-1}\nleftarrow):= & \{\alpha_{(k-1)k}\, |\, \exists\, \alpha_{(k-2)(k-1)}: A_{k-1}(\alpha_{(k-2)(k-1)},\alpha_{(k-1)k}) = 0 \}\\
\nonumber 
& \text{This is the set allowing $A_{k-1}$ to output $0$ for some of the hidden variables that come from the other side.} \\ 
\nonumber 
& \text{That is, \emph{classical photons are not sent to $A_{k-1}$ from the right.}}\\
\label{eq:setapp3}
(\rightarrow^k):= & \{\alpha_{(k-1)k}\, |\, \exists\, \alpha_{k(k+1)}: A_k(\alpha_{(k-1)k},\alpha_{k(k+1)}) = 2 \}\\
\nonumber 
& \text{This is the set allowing $A_k$ to output $2$ for some of the hidden variables that come from the other side.} \\
\nonumber 
& \text{That is, \emph{some classical photons are sent to $A_{k}$ from the left.}}\\
\label{eq:setapp4}
(^{k-1}\leftarrow):= & \{\alpha_{(k-1)k}\, |\, \exists\, \alpha_{(k-2)(k-1)}: A_{k-1}(\alpha_{(k-2)(k-1)},\alpha_{(k-1)k}) = 2 \}\\
\nonumber 
& \text{This is the set allowing $A_{k-1}$ to output $2$ for some of the hidden variables that come from the other side.} \\
\nonumber
& \text{That is, \emph{some classical photons are sent to $A_{k-1}$ from the right.}}
\end{align}

\subsection{Constraints on the sets}
We here derive in this generalized $N$-party scenario the constraints on any local model reproducing $p_t$ corresponding to those obtained for the triangle network (\ref{subsec:const1} to \ref{sec:all_sets_0.5}).

As depicted in Fig.~\ref{fig:n-parties} we have, firstly,
\begin{align}
(\nrightarrow^k)\cap(^{k-1}\nleftarrow)=\emptyset \;,
\end{align}
because otherwise two neighbouring parties $A_{k-1},A_k$, would be allowed to output $0$ at the same time, which is in contrast with the output of $p_t$ (the photon shared between two parties ends up in one of their detectors). 

Secondly
\begin{align}
(\nrightarrow^k)\cup(^{k-1}\nleftarrow)=1\;,
\end{align}
meaning that, together, the two sets form the total set of sources $\alpha_{(k-1)k}$ between $A_{k-1}$ and $A_k$. This is proven as a consequence of the fact that at least one between $A_{k-1}$ and $A_k$ must be allowed to output 0 (otherwise there would be a non-zero probability of more than $N$ photodetectors firing, as in $\{a_{k-1}=\chi,a_k=\chi, a_{k+1}=2,\chi, \chi, \chi \dots\}$). The initial total number of photons is $N$, therefore this cannot happen.

Thirdly we have
\begin{align}
(\rightarrow^k)\subseteq(^{k-1}\nleftarrow)\;, \\
(^{k-1}\leftarrow)\subseteq(\nrightarrow^k)\;.
\end{align}
This is true again because otherwise an event like $\{a_{k-1}=\chi,a_{k}= 2,a_{k+1}=\chi, \chi, \chi,\dots\}$ or $\{a_{k-1}=2,a_{k-1}=\chi,a_{k+1}= \chi, \chi, \chi,\dots\}$ would have nonzero probability. 
All the above constraints are derived out of photon number conservation (note that in our optical setup, if we do not resolve the number of photons, sometimes we may lose track of some of them when they end up in the same detector, which is why we are not able to say that the above equations are equalities, but just inclusions). 

Now, it is also true that
\begin{align}
\label{eq:sets_0.5}
\text{all the } (\nrightarrow^k) \text{ and } (^k\nleftarrow) \text{ sets have probability equal to } \frac{1}{2} \forall k\;.
\end{align}
This can be proven by using the definitions as
\begin{equation}
(\nrightarrow^k)*(^k\nleftarrow)\geq p(a_k=0)=\frac{1}{4}=\left(\frac{(^k\nleftarrow)+(\nrightarrow^{k+1})}{2}\right)^2\geq (^k\nleftarrow)*(\nrightarrow^{k+1})
\end{equation}
which implies $(\nrightarrow^k)\geq (\nrightarrow^{k+1})$, but such inequality can be cycled until obtaining $(\nrightarrow^k)\geq (\nrightarrow^{k})$, which entails that all the inequalities are actually equalities.

\subsection{Constraints on the local coarse grained strategy}
Given~\eqref{eq:sets_0.5}, we have that the parties will output deterministically $0$ when allowed from both sides, as they have to simulate $p(0)=\frac{1}{4}$. Summing up we have
\begin{align}
\label{eq:propapp0}
(\nrightarrow^k) A_k (^k\nleftarrow) & \quad \Rightarrow  A_k \text{ outputs } 0 \\
(^{k-1}\nleftarrow) A_k (^k\nleftarrow) \quad \text{ or } \quad (\nrightarrow^k) A_k (\nrightarrow^{k+1}) & \quad \Rightarrow A_k \text{ outputs } \chi \\
(^{k-1}\nleftarrow) A_k (\nrightarrow^{k+1}) &  \quad \Rightarrow A_k \text{ outputs } \chi \text{ or } 2 
\end{align}

\subsection{Breaking the coarse-graining and finding linear constraints}
Here we repeat and generalize the scheme presented in~\ref{sec:breaking_up} to give linear constraints on a subset of the local response functions. We define $q_t(i_1,i_2,...,i_N,s)$, analogously to \eqref{eq:q_DEF} as the probability of outputting $\{\chi_{i_1},\chi_{i_2},...,\chi_{i_N},\}$ given sources $\alpha$s coming from left ($s=0$) or right ($s=1$) part of the sets drawn in Fig.~\ref{fig:n-parties}, i.e.
\begin{align}
q_t(i_1,i_2,...,i_N,s)=
\begin{cases}
p_t(a_1=\chi_{i_1},a_2=\chi_{i_2},...,a_N=\chi_{i_N},(\alpha_{12},\alpha_{23},\dots)\in S_0) & s=0\\
p_t(a_1=\chi_{i_1},a_2=\chi_{i_2},...,a_N=\chi_{i_N},(\alpha_{12},\alpha_{23},\dots)\in S_1) & s=1
\end{cases}
\end{align}
where we formally define the above mentioned sets as 
\begin{align}
\nonumber
    S_0:={\times_j} (\nrightarrow^j)\equiv (\nrightarrow^1)\times (\nrightarrow^2)\times\dots\times(\nrightarrow^N)\;,\\
    S_1:={\times_j} (^j\nleftarrow)\equiv(^1\nleftarrow)\times(^2\nleftarrow)\times\dots\times(^N\nleftarrow)\;.
\end{align}
Given that the configurations of sources ${\times_j} (\nrightarrow^j)$ and  ${\times_j} (^j\nleftarrow)$ are the only ones allowing possible outputs being all $\chi$, $q$ satisfies the following equality involving one of its marginal distributions
\begin{align}
\label{eq:marg_app_1}
\sum_{s=0,1} q_t(i_1,i_2,...,i_N,t)= p_t(\chi_{i_1},\chi_{i_2},...,\chi_{i_N})\;.
\end{align}
We now consider instead the marginal on $i_k$ and $t$
\begin{align*}
q_t(i_k,s)=\sum_{i_1,...,i_{k-1},i_{k+1},...,i_N} q_t(i_1,i_2,...,i_N,s)\;.
\end{align*}
This satisfies
\begin{align}
\label{eq:marginal_n-parties}
q_t(i_k,0)-q_t(i_k,1)=\frac{p_t(a_k=\chi_i,a_{k+1}=0)-p_t(a_{k-1}=0,a_k=\chi_i)}{2^{N-3}}\;.
\end{align}
The proof of this equation is formalized as follows
\begin{multline}
q_t(i_k,0)-q_t(i_k,1)\\
=p_t(a_k=\chi_{i_k},{\times_j} (\nrightarrow^j))-p_t(a_k=\chi_{i_k},{\times_j} (^j\nleftarrow))\\
=p_t\left(a_k=\chi_{i_k},({\times_{j\neq k+2}} (\nrightarrow^j))\times (^{k+1}\nleftarrow) \right) - p_t\left(a_k=\chi_{i_k},({\times_{j\neq k-2}} (^j\nleftarrow))\times (\nrightarrow^{k-1})\right)\\
=\frac{1}{2^{N-3}}\left[
p_t\left(a_k=\chi_{i_k},(\nrightarrow^k)\times(\nrightarrow^{k+1})\times(^{k+1}\nleftarrow) \right) - p_t\left(a_k=\chi_{i_k},(\nrightarrow^{k-1})\times(^{k-1}\nleftarrow)\times(^k\nleftarrow)\right)\right]\\
=\frac{p_t(a_k=\chi_i,a_{k+1}=0)-p_t(a_{k-1}=0,a_k=\chi_i)}{2^{N-3}}\;.
\end{multline}
The first equality above simply follows from the definition of $q_t(i_k,s)$ for $s=0,1$.
The second equality is obtained by noticing that all sets $(\nrightarrow^j)$ and $(^j\nleftarrow)$ have probability $1/2$, and that the output $a_k$ does not depend on the source shared between $A_{k+1}$ and $A_{k+2}$, nor it depends on the source shared between $A_{k-2}$ and $A_{k-1}$.
The third equality is obtained by tracing out the probability of $N-3$ of the sets which were included in the previous lines.
Finally the last inequality is implied by property \eqref{eq:propapp0}.

The above constraints on $q_t$ coincide with the ones derived in the Appendix~C of \cite{renou2019genuine}. There, it is proven that it is always possible to choose the value of the transmissivity $t$ such that no solution can be found for $q_t(i_1,i_2,\dots,i_N,s)$ satisfying the linear constraints \eqref{eq:marg_app_1} and \eqref{eq:marginal_n-parties}.
Therefore for those values $t$ the output $p_t$ of the experiment is proven to be nonlocal.


\end{document}